\documentstyle[aps,preprint,eqsecnum,floats,epsf,rotate]{revtex} 
\tighten
\draft
\begin{document}

\title{Supernova Neutrino-Nucleus Astrophysics}
\author{A.~B. Balantekin\thanks{Electronic address:
        {\tt baha@nucth.physics.wisc.edu}}}
\address{Department of Physics, University of Wisconsin\\
         Madison, Wisconsin 53706 USA}
\author{G.~M. Fuller\thanks{Electronic address:
        {\tt gfuller@ucsd.edu}}}
\address{Department of Physics, University of California, San Diego\\
         La Jolla, CA 92093-0319 USA}

\maketitle

\begin{abstract}
In this brief review we explore the role of neutrino-nucleus 
interactions in core-collapse supernovae and discuss open 
questions. In addition implications of neutrino mass and mixings 
in such environments are summarized.

\end{abstract}

\date{\today} 
 
\newpage

\section{Open Questions for Neutrino Interactions in Core-collapse
   Supernovae}

We do not as yet understand how supernovae explode or even  whether
explosion is a common outcome of a core collapse event. The  current
\lq\lq nuclear physicist's paradigm\rq\rq\ (that is,
leaving out rotation and
magnetic fields) for explosions is as follows: a massive star evolves
on a timescale of millions of years and forms a Chandrasekhar mass
core composed of iron peak nuclei. The weak interaction dominates the
evolution of an object like this and neutrino emission from core
carbon/oxygen burning onward efficiently removes entropy from the
star,  ultimately causing the iron core to have a very low entropy per
baryon. As a result, the core is supported by relativistically
degenerate electrons and this fact has immediate implications: the
core will go dynamically unstable and will collapse at a near free
fall rate on a very low adiabat until the nuclei and nucleons merge at
nuclear density. A shock wave is generated at this point,
essentially at the edge
of an inner homologous core which is a sort of instantaneous
Chandrasekhar mass. During the collapse,
the electron Fermi energy rises
as the volume of the core decreases. This
drives copious electron capture
on protons, lowering the fraction of electrons per baryon, $Y_e$,
and, hence, lowering the instantaneous Chandrasekhar mass. The protons
which are the targets for electron capture are mostly
inside large nuclei on account
of the low entropy. The physics of this process is discussed in detail
in Ref. \cite{Langanke:2003ii}.

At issue is
the fate of this \lq\lq bounce\rq\rq\ shock (see e.g. Ref. 
\cite{Blondin:2002sm}). The mere fact that
nucleons are bound in nuclei by some 8 MeV on average implies that
this initial  shock will be \lq\lq dead on arrival.\rq\rq\ As the
shock transits material beyond the inner core, most
of it's kinetic energy
is dissipated in the photo-dissociation of nuclei. The
shock quickly ($\sim 100\,{\rm milliseconds}$) evolves to become a
standing  accretion shock.

This process has to happen for two reasons: (1) a strong shock will
have a large (by nuclear physics standards) jump in entropy  across it
($\sim$ factor of ten in this case); and (2) the temperature is high
enough that all the material is in nuclear statistical
equilibrium. All nuclei will be \lq\lq melted\rq\rq\ if the entropy is
increased by more than three or four units of Boltzmann's constant per
baryon.

During the collapse of the core, the infall epoch, electrons are
captured on protons which for the most part, reside in nuclei. The
electron neutrinos produced in this process freely escape from the
core at first, but later become trapped at densities exceeding one
percent of nuclear matter density. Subsequently, the electron
neutrinos scatter and exchange energy and approach beta equilibrium. 
However, the manner in which the neutrinos approach equilibrium may
depend on neutrino-nucleus interaction rates. These include processes
in which nuclear excitation energy is changed into neutrino energy and
processes in which neutrinos give up energy to nuclei
\cite{oldmeyer}. For example, it could be that the de-excitation of
hot nuclei into neutrino-antineutrino pairs is the dominant source of
low-energy neutrinos and the principal means by which the low-energy
neutrinos are driven into equilibrium during the infall epoch. This
needs to be investigated further. 

The gravitational binding energy release in the collapse to nuclear
density, some one percent of the core rest mass promptly and,
ultimately, ten percent of the rest mass, is radiated as neutrinos of
all flavors. It is believed that some of this neutrino  energy is
transferred to thermal energy behind the otherwise stalled shock on a
timescale of some 100's of milliseconds post bounce. How the neutrino
energy can be transported to and pumped into this region in and around
the shock is still very much an open question.   Neutrino-nucleus
interactions are expected to play an important role.

Electron neutrino captures on neutrons (making protons) and electron
antineutrino captures on protons (making neutrons) will proceed apace
in this region. This is likely the dominant way in which energy can be
transferred to the shock from the neutron star directly via
neutrinos. (Convection through the neutrinosphere can be regarded as
an effective increase in neutrino luminosity. Convection between the
neutrinosphere and the shock is different, and may be quite important for
obtaining an explosion, and will be ignored here.) If the core
(proto neutron star) is highly relativistic, then neutrino-antineutrino
annihilation may play a significant role in energy deposition as
well. These charged current captures on free nucleons also determine
the neutron to proton ratio in the material above the neutron star.
The neutron excess of the neutron star is \lq\lq transmitted\rq\rq\
to the overlaying material via neutrinos and these charged current
capture processes.

Electron neutrino and antineutrino captures on heavy nuclei ahead of
the shock may be important. If enough energy can be transferred to
these nuclei from these processes to melt a fraction of the nuclei
then the shock dynamics can be altered. This \lq\lq pre-heating\rq\rq\
of the material ahead of the
shock (a misnomer as the material has a large specific heat on account
of the heavy nuclei
and the temperature does not rise much)
could help the shock by partially dissociating
nuclei; and, it could hurt the shock by melting nuclei, giving a
higher number density of particles and therefore producing a higher
pressure. The Mach number of the shock goes roughly like the pressure
ratio across the shock front and it is obvious that melting the nuclei
ahead of the shock goes in the 
direction of evening up the pressure on both sides.  Which of
these effects wins is not at all clear at present. To predict the
outcome with confidence we need a better handle on neutrino-nucleus
interaction cross sections, among other things.

In general, the cross sections per nucleon for these processes are
small compared to those for free nucleons. However, collectivity in
the electron neutrino charged current channel (for nuclei with a
neutron excess) and in the (all neutrino flavors) neutral current
channel can result in significant cross sections that may be
important. To this end one needs neutral- and charged-current
neutrino-nucleus inclusive cross section for neutrino energies
ranging from 10's of MeV to 100 MeV or so,  principally for the iron
peak nuclei.

Another way in which these processes can be interesting is in the
freeze-out nucleosynthesis that may result from this shock re-heating
epoch. Again, the neutron to proton ratio will be set by the charged
current captures on free nucleons, but neutrino-nucleus reactions can
be important.

For example, electron neutrino capture on a neutron rich
target nucleus will tend to leave the daughter nucleus in a highly
excited state. The Gamow-Teller strength distribution centroid will
tend to lie in
the vicinity of or above the excitation energy of the first isobaric
analog state in the daughter. In Figure 1
we summarize the systematics of the
nuclear physics of neutrino
capture on a neutron rich target
nucleus. The essential point of physics is this:
the Coulomb energy sets the scale for
daughter nucleus excitation energy.
This can be $\sim 30\,{\rm MeV}$ or more
for a target (parent) like $^{208}$Pb.
Note that such a highly excited (massive)
daughter nucleus could decay by emission of one
or more neutrons, or even by fission. The
decay of such neutrino capture-produced
nuclei needs to be better
understood, especially as regards
the branching ratio into one or more
neutrons and the distribution
of fission fragments with mass.
This in turn may shed light on, for example, where  the
light p-process nuclei, like $^{96}$Mo, come from.

\begin{figure}
\centering\leavevmode
\rotate[r]{\epsfxsize=5.1in
\epsfysize=6.5in
\epsfbox{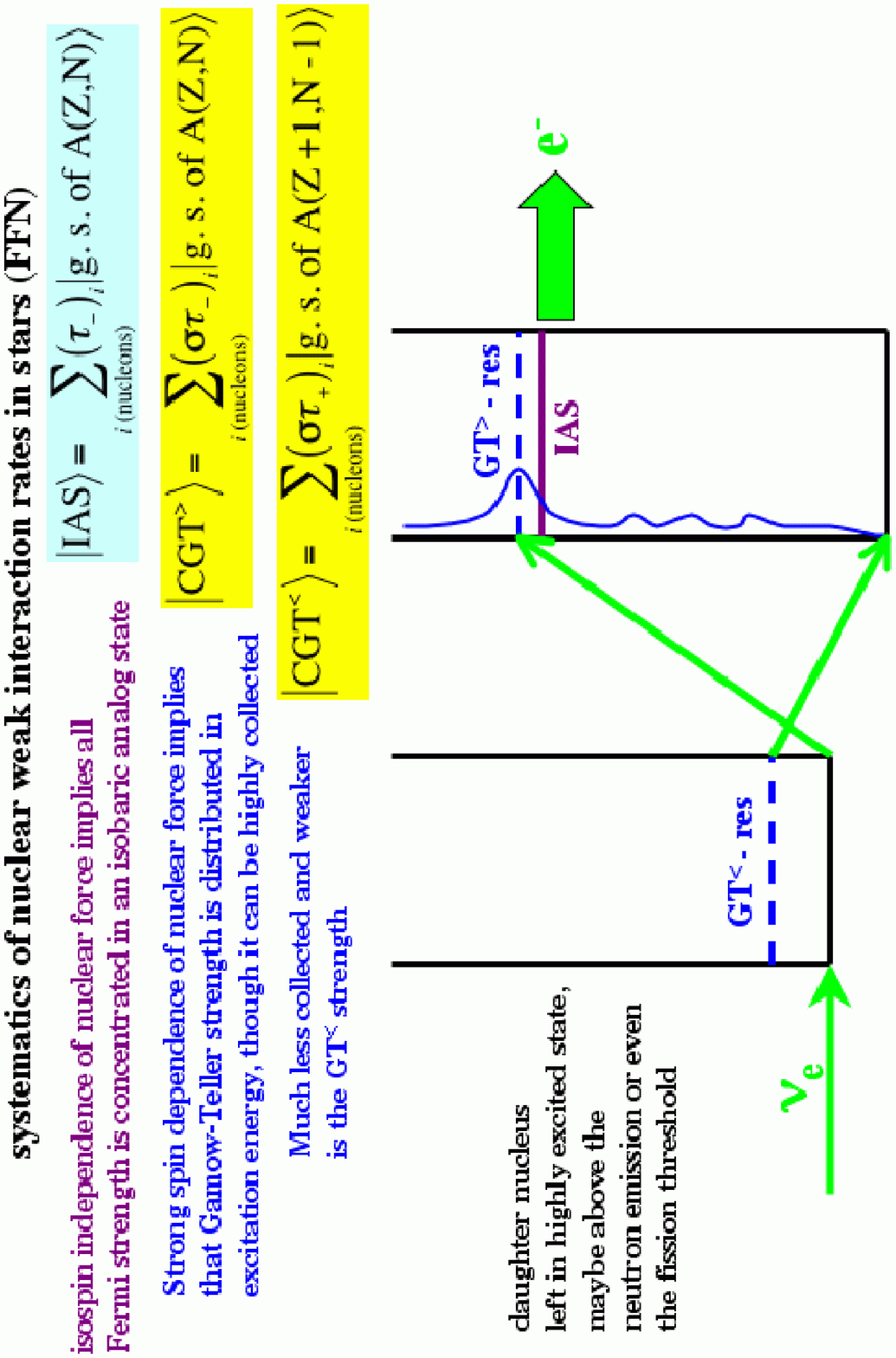}}
\caption{The systematics of the nuclear physics of neutrino
capture on a neutron rich target nucleus.}
\end{figure}

If there is an explosion and the material overlying the neutron star
is ejected, or if we had a non-explosive accretion-driven collapse
event, then we could be left with conditions conducive to the
formation of a neutrino-driven wind. Intense neutrino fluxes from the
neutron star can deposit energy (as outlined above) in the tenuous
medium above the neutron star, heating it to high entropy
and driving a wind. The entropy per baryon here
might be some hundreds
of units of Boltzmann's constant per baryon, which is
indeed high from the standpoint of
nuclear physics, as nuclei beyond alpha 
particles would be greatly disfavored 
in conditions of nuclear statistical equilibrium.

In any case, it is revealing to consider the gross energetics
of the neutrino-driven wind, at least insofar as the ejection
of baryonic material is concerned. The gravitational binding energy
of a nucleon near the neutron star surface will be $\sim 100\,{\rm MeV}$.
To be ejected into interstellar space, this baryon will have to acquire
an equivalent amount of energy from heating processes, either
transferred directly via weak interactions (neutrino captures
or neutrino-antineutrino annihilation) or via hydrodynamic
or convective transport of neutrino energy from the core
(see the Mezzacappa contribution in this volume). If the energy
for ejection is deposited by neutrino capture, and since average
neutrino energies are some $\sim 10\,{\rm MeV}$, we would require
a flux of electron neutrinos
and antineutrinos sufficient that each baryon ejected interact on
average some 10 times with neutrinos. Given the strength
of the weak interaction, it is a truly prodigious flux
of neutrinos which is required.

\begin{figure}
\centering\leavevmode
\rotate[r]{\epsfxsize=5.5in
\epsfysize=6.5in
\epsfbox{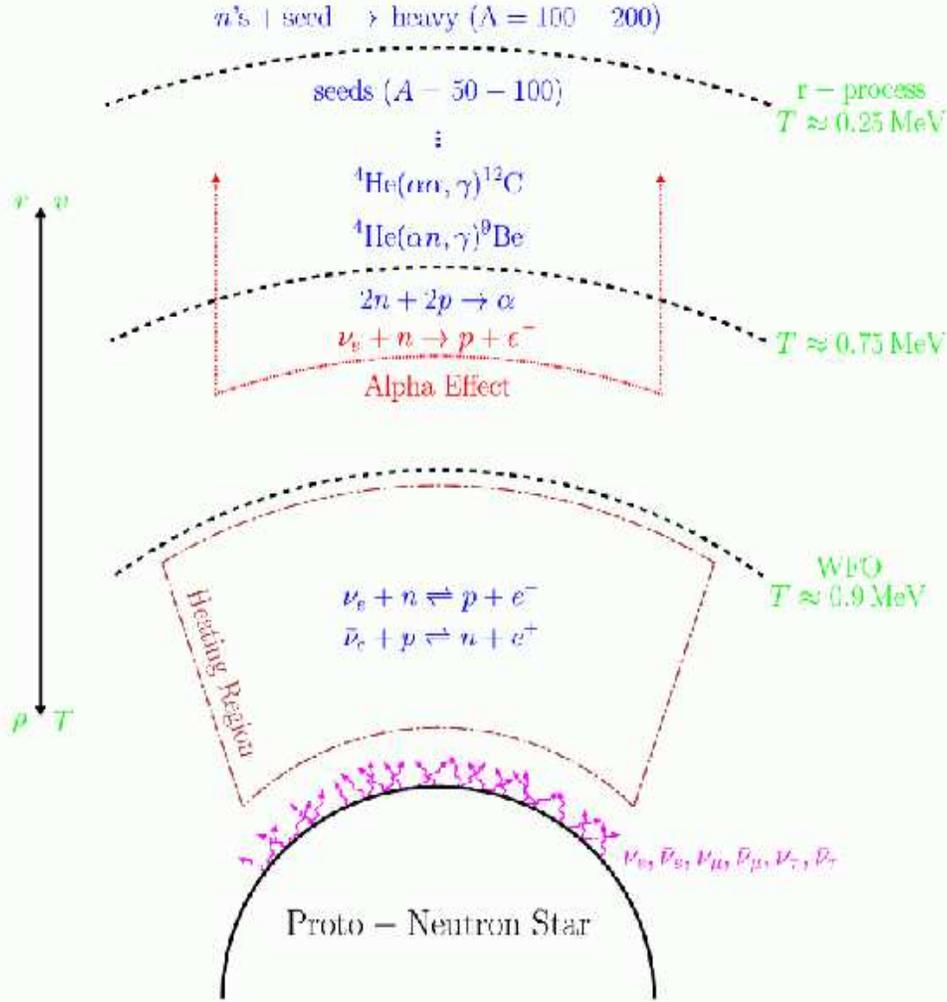}}
\caption{Outflow and nucleosynthesis history
of a slow expansion neutrino-driven wind in a core-collapse 
supernova.}
\end{figure}

In Figure 2 we show a cartoon of the outflow and nucleosynthesis history
of a \lq\lq slow expansion\rq\rq\ neutrino-driven wind.
Overall, the bulk of this \lq\lq wind\rq\rq\ is really
a near hydrostatic (subsonic material transport speeds),
near constant specific entropy envelope with a small
total mass ($\sim {10}^{-4}\,{\rm M}_\odot$).
With the approximation that the entropy
is mostly carried by relativistic particles (photons
and electron/positron pairs), we can relate the
temperature in billions of Kelvins to the radius in units
of ${10}^6\,{\rm cm}$ and the entropy in units of
$100$ Boltzmann's constant per baryon, $r_6 \approx
{{22.5}\over{T_9 s_{100}}}$,
implying that the overall mass density falls as the
inverse cube of the radius.
Neutrino driven wind models, where the outflow is homologous
(i.e. fluid velocity is proportional to the distance) are
characterized by two parameters: entropy per baryon and the expansion
timescale, $\tau$ (i.e. $r=r_0 e^{t/\tau}$, where the expansion rate
of the baryonic material is $\lambda_{\rm exp} \equiv 1/\tau$).

Conventional slow expansion wind models are characterized by the
establishment of steady state weak equilibrium and by several alpha
particle formation-related problems in the production
of the r-process nuclides (see below). However, if there was
a way around these alpha particle issues, then the slow expansion
wind could in principle produce an r-process (though
absent active-sterile matter-enhanced neutrino flavor
conversion this is difficult).

By contrast, rapid outflow, perhaps occasioned by a very compact
proton neutron star with consequent large general relativistic effects 
\cite{Cardall:1997bi,Thompson:2001ys} 
can circumvent the
the various alpha particle/neutrino flux related problems with the
r-process, but perhaps at the cost of producing either
not enough r-process material or a distribution of r-process
abundances which look nothing like the ones we observe in the solar
system or in ultra metal-poor halo stars.

Close in 
to the neutron star surface, neutrino capture reactions on free nucleons
transfer energy and heat the medium to high entropy. The competition
between electron antineutrino capture on protons and electron neutrino
capture on neutrons sets the neutron to proton ratio. The Weak Freeze
Out position (WFO) occurs where the rates of these isospin changing
reactions falls below the expansion rate of the baryonic flow
(roughly $T_9 \approx 10$). Further
out where it is cooler, the bulk of charged nuclear reactions
become slow compared to the expansion rate, which we
can term NSE freeze out
(the crude equivalent of the alpha particle formation epoch in BBN).
Still further out, neutrons can capture on the relatively few seed
iron nuclei produced near NSE freeze out.

The nucleosynthesis in
the freeze-out of this ejected material might closely resemble Big Bang
Nucleosynthesis (BBN), albeit with a neutron excess instead of the
proton excess which obtains in the early universe! However, unlike
BBN, this near isospin mirror of BBN could produce rapid neutron capture
(r-process) nucleosynthesis. Neutrons could capture on iron peak seed
nuclei and yield heavy nuclei, like uranium.

All of this neutron
capture will be proceeding in an environment with large fluxes of
neutrinos. As we discuss below,
the formation of the alpha particles in a region
where the neutrino fluxes are still high can lead to a fatal
problem for this scenario for the r-process.
In any case, neutrino flavor transformation at the
atmospheric neutrino mass-squared difference scale will likely produce
an energetic spectrum of electron neutrinos. Recently, it has been
shown that the electron neutrino capture-induced fission cross
sections for heavy nuclei can be very large (e.g.,
approaching $10^{-38}\,{\rm
cm}^2$ for the actinides) \cite{Qian:2002mb,Kolbe:2003jf}.
Again, neutrino capture in this channel tends to leave the
daughter nucleus in a highly excited state which, in a heavy nucleus,
may be well above the fission barrier.  One needs to know the
branching ratios into multiple neutron emission and into fission, and
if the excited daughter nucleus fissions, then we need
to know the typical number of neutrons
that come off.

One needs to ascertain whether or not the fission rate stemming  from
electron neutrino captures can be big enough to drive fission cycling
in the r-process flow. This seems unlikely, however, given
the large numbers of neutrons per fission fragment 
that would
be required to build these fragments back to massive nuclei.
To this end, we would like to know  what the
branch into fission for heavy nuclei and the distribution of fission
fragment masses are. Capture on nuclei 
in the mass 195 peak, followed by fission, will likely give fragments
with masses near or in the 130 peak. This may tie the abundances in
these peaks together. The astronomical observations of ultra
metal-poor halo stars suggests that there is a physical connection
between these mass regions and the processes which produce them 
\cite{Sneden:2003zq}.

\section{Implications of Neutrino Mixing in Supernovae}

Matter-enhanced neutrino flavor transformations, in both the
active-active and active-sterile channels, can have a significant
effect on the dynamics, nucleosynthesis, and the neutrino signal
associated with core-collapse supernova events. Here we emphasize the
active-sterile channel because that is most likely to have a dramatic
effect on the outcome of the stellar collapse. We ignore the rich
physics that can be explored by the terrestarial detection of these
neutrinos \cite{Beacom:2002hs,Haxton:kc,Schirato:2002tg}. 
Experimental and observational evidence for a new neutrino 
mass-squared
difference ($\delta m^2$) scale, different from the atmospheric
neutrino scale ($\delta m^2 \sim 3  \times{10}^{-3}\,{\rm eV}^2$)
and the solar neutrino value ($\delta  m^2 \sim 7 \times {10}^{-5} 
\,{\rm eV}^2$)
would likely provide the smoking gun for the existence of sterile
neutrinos. For example, 
the $\delta m^2$ range being explored currently by the mini-BooNE
experiment covers an important range in the
$\nu_{\mu,\tau}\rightleftharpoons\nu_e$ channel, that bears on the
question of whether light sterile neutrinos exist. 
It has recently been shown that 
matter-enhancement of 
$\nu_e\rightleftharpoons\nu_s$, 
$\bar{\nu_e} \rightleftharpoons \bar{\nu_s}$ 
in the supernova can provide an ideal environment for rapid
neutron capture (r-process) nucleosynthesis and solve
a fundamental problem in the slow outflow scenario r-process
alluded to above.

The importance that experiments
like mini-BooNE have flow from the increase in  reliability and
precision in the identification and measurement of  neutrino
oscillation phenomena in atmospheric neutrinos
\cite{Fukuda:1998ub,Fukuda:1998tw,Fukuda:2000np}
and solar neutrinos at Sudbury Neutrino Observatory
\cite{Ahmad:2002jz,Ahmad:2001an}, SuperKamiokande
\cite{Fukuda:2001nk}, Chlorine \cite{Cleveland:nv}, and Gallium
\cite{Abdurashitov:2002nt,Hampel:1998xg,Altmann:2000ft} experiments.
These experiments are  so restrictive in their
definition of the allowed neutrino mixing  parameter space that there
is \lq\lq no room\rq\rq\ for mixing at an  additional $\delta m^2$
scale. This result is confirmed by many independent analyses
\cite{Balantekin:2003dc,allothers} which also take into account the
recent data from the KamLAND reactor neutrino experiment
\cite{Eguchi:2002dm}.
The Los Alamos Liquid Scintillator Neutrino Detection (LSND)
experiment has reported an excess of $\bar\nu_e$-induced events above
known backgrounds in a $\bar\nu_\mu$ beam with a statistical
significance of $3$ to $4$ $\sigma$
\cite{Athanassopoulos:1996jb,Aguilar:2001ty}.
If this result is confirmed by mini-BooNE it
represents just such evidence for vacuum neutrino oscillation at a new
$\delta m^2$ scale.  Discovery of such mixing
would imply either CPT-violation  in the
neutrino sector,
or the existence of a light singlet  \lq\lq sterile\rq\rq\
neutrino which mixes with active species. The latter explanation may
signal the presence of a large and unexpected net lepton number in the
universe and the existence of a light singlet complicates the
extraction of a neutrino mass limit from Large Scale Structure data.
Either explanation could alter our models for core collapse supernova
explosions and the origin of heavy elements in neutrino-heated
supernova ejecta. Indeed, r-process abundance observations and
calculations by themselves hint that the mass-squared difference range
$0.1\,{\rm eV}^2 < \delta m^2 < 100\,{\rm eV}^2$ in the channel
$\nu_{\mu,\tau}\rightleftharpoons\nu_e$ is worthy of experimental
exploration.

R-process nucleosynthesis
constrained to reproduce something like the solar
system abundance pattern
(which the ultra metal-poor, or UMP,
halo star data indicates may be universal)
requires a neutron-rich environment, i.e.,
the ratio of electrons to baryons, $Y_e$, should be less than one
half. Arguments based on meteoritic data
and the systematics of abundances in UMP's suggests that one
possible site for r-process nucleosynthesis is the neutron rich
material associated with core-collapse supernovae
\cite{Qian:1998,Qian:1998cz}.  In outflow models, freeze-out from
nuclear statistical equilibrium leads to the r-process
nucleosynthesis. The outcome of the freeze-out process in turn is
determined by the neutron to seed ratio. The neutron to seed ratio is
controlled by three quantities: i) The expansion rate; ii) The
neutron to proton ratio (or equivalently the electron fraction,
$Y_e$); iii) The entropy per baryon. Of these three the
neutron to proton ratio is completely determined by the
neutrino-nucleon and neutrino-nucleus interactions.

Crudely, the electron fraction in the
nucleosynthesis region is given approximately
by \cite{Qian:dg}
\begin{equation}
Y_e \simeq {1 \over 1+ \lambda_{{\overline \nu}_e p} / \lambda_{ \nu_e
n}}  \simeq {1 \over 1 + T_{{\overline \nu}_e} / T_{ \nu_e}},
\end{equation}
where $\lambda_{ \nu_e n}$, etc. are the capture rates and various
neutrino temperatures are indicated by $T$.
This expression ignores the possibility
that the luminosities of neutrinos of
different flavors are different and it ignores weak magnetism
corrections, for example. Note that weak magnetism corrections in this
case go in the direction of decreasing neutron excess and therefore
increasing the difficulty of obtaining a viable r-process 
\cite{Horowitz:2003yx}. 
Hence if $T_{{\overline
\nu}_e} > T_{\nu_e}$, then the medium is neutron rich.  Without
matter-enhanced neutrino oscillations, the neutrino temperatures
in some models
satisfy the inequality $T_{ \nu_{\tau}} >T_{{\overline \nu}_e} > T_{
\nu_e}$. But matter effects via the MSW mechanism
\cite{Balantekin:1998yb}, by heating $\nu_e$ and cooling $\nu_{\tau}$,
can reverse the direction of inequality, making the medium proton rich
instead. Hence the existence of neutrino mass and mixings puts
an interesting twist on the production of
heavy-elements in supernovae. These
connections are investigated in Refs. \cite{Qian:dg} and
\cite{Qian:wh}. One should also point  out that in stochastic media
(i.e. media with large density fluctuations) neutrino flavors would
depolarize \cite{Loreti:1994ry,Balantekin:1996pp}. Although recent
solar neutrino experiments rule out such effects for the Sun
\cite{Balantekin:2003qm}, they may be important in supernovae.
\cite{Loreti:1995ae}.

It could also be (and this seems increasingly likely) that neutrino
opacity sources not previously taken into account essentially wipe out the
hierarchical average neutrino energy picture described above. 
If this is
true, then all neutrino species would possess roughly
the same energy spectrum. If the luminosities of the different
neutrino flavors are the same, then we can draw two conclusions
for this case: (1) the conditions are not neutron
rich (in fact the neutron to
proton ratio would be close to unity) in a slow outflow; and
(2) active-active matter-enhanced neutrino flavor conversion
would have essentially no effect. This clearly
would exacerbate the other problems with obtaining
a solar system-like abundance pattern in a slow out flow
neutrino-driven wind.

These other problems
include, for example,
the so called \lq\lq alpha effect.\rq\rq\
To understand this, note that
there are two kinds of neutrino reactions that can destroy the
r-process scenario outlined above:
i) neutrino neutral current spallation of alpha particles
\cite{meyer}; ii) formation of too many alpha particles
in the presence of a strong
electron neutrino flux, known as the
alpha effect \cite{Fuller:ih,Meyer:1998sn}. The alpha effect comes
at the epoch of alpha-particle formation: protons produced by $\nu_e$
capture on neutrons will, in turn, capture more neutrons to bind into
alpha particles, reducing the number of free neutrons available to the
r-process and pushing $Y_e$ towards $0.5$. Reducing the $\nu_e$ flux
will resolve this problem, but we can only do so at a relatively large
radius so that effective neutrino heating already can have
occurred. One way to achieve this is transforming active electron
neutrinos into sterile neutrinos
\cite{McLaughlin:1999pd,Caldwell:1999zk,Patel:1999hm,fetter}.

For active-sterile neutrino mixing
in the channels $\nu_e \rightleftharpoons \nu_s$
and ${\bar\nu}_e \rightleftharpoons {\bar\nu}_s$, and
for $Y_e > 1/3$, only electron
neutrinos, and for $Y_e < 1/3$ only electron antineutrinos can undergo
an MSW resonance \cite{McLaughlin:1999pd}. If both electron neutrino
and antineutrino fluxes go through a region where
the isospin is in steady state weak
equilibrium (i.e. the reactions
$\nu_e + n \rightleftharpoons p + e^-$ and
$\bar{\nu}_e + p \rightleftharpoons n + e^+$
are in steady state equilibrium
with the $\nu_e$ and $\bar\nu_e$ fluxes), then no matter what the
initial $Y_e$ is one may expect that the system will evolve to a fixed
point with $Y_e = 1/3$ ensuring a neutron rich medium
\cite{Nunokawa:1997ct}. Realistic calculations of the supernova wind
models \cite{McLaughlin:1999pd} do not bear out this assessment;
although the electron antineutrinos are converted into sterile
species, they are regenerated before the electron fraction
and the isospin in the wind
freeze out of steady state equilibrium.

\begin{figure}
\centering\leavevmode
\rotate[r]{\epsfxsize=2.5in
\epsfysize=3.5in
\epsfbox{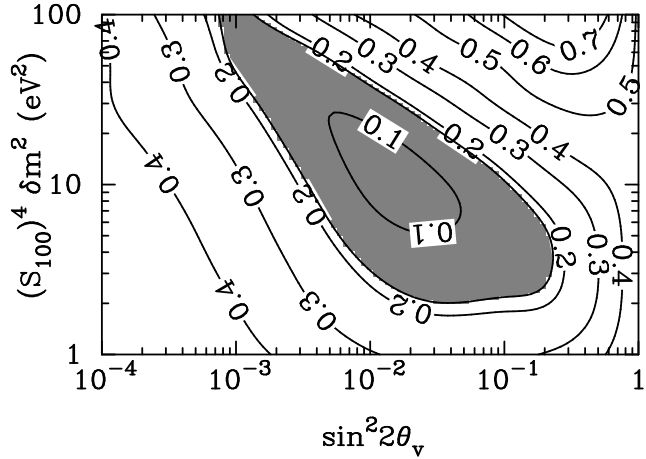}}
\caption{Contours of electron fraction for a timescale of 0.3 s in the 
active to sterile conversion scenerio. The shaded area yields a neutron 
to seed ration of at least 100. $Y_e \sim 0.5$ in both with no flavor 
transformation.}
\end{figure}

In Refs.
\cite{McLaughlin:1999pd} and \cite{fetter} we followed the neutrino
flavor evolution equations in the wind
in a manner which was self consistent with
the neutrino capture reactions which set the
neutron to proton ratio and, hence, $Y_e$. Additionally, we tracked the
thermodynamic and nuclear statistical equilibrium evolution of
outflowing mass
elements and updated the isospin of these at each time
step directly from the weak capture rates.  This coupling of the
neutrino evolution and self-consistent determination of the abundances
is essential to accurately determine the number of neutrons available
for the r-process. The results are illustrated in Figures 3 and 4 for
expansion timescales of $\tau=0.3$ and $\tau=0.9$ seconds
respectively. It can be concluded from these figures
that there is a wide range of neutrino mass/mixing
parameters (some even consistent
with the LSND parameters when we take account
of the dependence of the results on
the uncertain entropy of the neutrino-driven wind)
which vitiates the alpha effect and can greatly enhance the
neutron-to-seed ratio to produce favorable conditions for r-process
nucleosynthesis. We note that values of the effective two-by-two
active-sterile mixing angle up to an order of magnitude smaller
than those probed by LSND ({\it i.e.}, smaller than those probe
by mini-BooNE) still give efficient flavor conversion in the hot bubble
wind environment of the supernova.

\begin{figure}
\centering\leavevmode
\rotate[r]{\epsfxsize=2.5in
\epsfysize=3.5in
\epsfbox{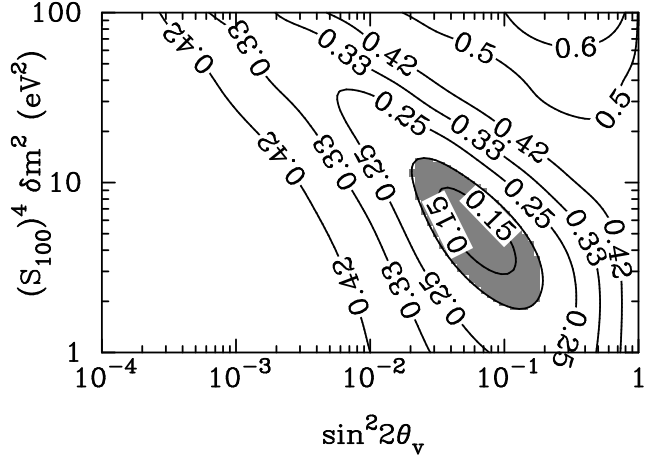}}
\caption{Same as Figure 3, but with a timescale of 0.9 seconds.}
\end{figure}

In fact, the electron fractions produced in these scenarios
involving matter-enhanced active-sterile neutrino flavor transformation
can be extremely small, perhaps with $Y_e < 0.1$. This is intriguing.
With neutron excesses this large it is easy to get a viable r-process.
Furthermore, a large enough neutron excess could cause fission-cycling
in the r-process flow, where neutron-rich nuclides capture enough neutrons
to become unstable against fission, and where, subsequently, the fission
fragments so produced themselves quickly capture enough neutrons to build
back to the nuclear mass where fission likely sets in, so that
a steady state flow results. Whether or not
neutrino captures or neutral current interactions could influence this
flow remains unclear at present but is a focus of ongoing
investigation. If the fissioning nuclides are at or above the
mass $A=195$ peak in the r-process, then the fission fragments could
be expected to be in the mass range of the
$A=130$ peak. This might help
explain an as yet poorly understood feature of the UMP halo
star r-process data: the fact that the total abundances in these
different nuclear mass peaks are about the same.

In any case, the connection between the physics of supernova dynamics
and heavy element nucleosynthesis on the one hand, and the
physics of neutrino mass and mixing on the other, remains a promising venue
for research.

We thank M. Patel for Figure 2, G. McLaughlin and J. Fetter for
Figures 3 and 4, A. Garcia for useful discussions. 
This work was supported in part by the U.S. National Science
Foundation Grant No.\ PHY-0244384  at the University of Wisconsin and
the U.S. National Science
Foundation Grant No. PHY-0099499 and DOE/SCIDAC Supernova Initiative 
at UCSD, and in part by the University of Wisconsin
Research
Committee with funds granted by the Wisconsin Alumni Research
Foundation.

\end{document}